\documentclass[aps,prl,showpacs,twocolumn,superscriptaddress]{revtex4}
\usepackage{bm,color}
\usepackage{graphicx}
\usepackage{amsmath}

\begin{document}
\title {Creation of Skyrmions by Electric Field on Chiral-Lattice Magnetic Insulators}

\author{Masahito Mochizuki}
\email{mochizuki@phys.aoyama.ac.jp}
\affiliation{Department of Physics and Mathematics, Aoyama Gakuin University, Sagamihara, Kanagawa 229-8558, Japan}
\affiliation{PRESTO, Japan Science and Technology Agency, Kawaguchi, Saitama 332-0012, Japan}

\begin{abstract}
We theoretically propose that magnetic skyrmion, nanometric spin vortex characterized by a quantized topological number, can be electrically created on a thin-film specimen of chiral-lattice magnetic insulator within a few nanoseconds by applying an electric field via an electrode tip taking advantage of coupling between noncollinear skyrmion spins and electric polarizations. This finding will pave a route to utilizing multiferroic skyrmions as information carriers for low-energy-consuming magnetic storage devices without Joule-heating energy losses.
\end{abstract}
\maketitle
Skyrmion was theoretically proposed by Tony Skyrme in 1962 as a topological soliton solution for the nonlinear sigma model to account for stability of baryons in the particle physics~\cite{Skyrme62}. Nowadays, skyrmions are attracting revived research interest in the community of condensed-matter physics. This revival started with theoretical predictions~\cite{Bogdanov89,Rossler06} and experimental observations~\cite{Muhlbauer09,YuXZ10N} of skyrmions as vortex-like topological spin textures in ferromagnets with chiral crystal symmetry.

In the chiral-lattice ferromagnets, Dzyaloshinskii-Moriya interactions (DMI) become active due to the broken spatial inversion symmetry and favors a rotating spin alignment,
while ferromagnetic-exchange interactions favor a parallel spin alignment. It was theoretically predicted that keen competition between these two interactions under a static magnetic field results in the formation of skyrmions as swirling spin textures and skyrmion crystals as triangular arrays of skyrmions. The skyrmion crystal was indeed observed in metallic chiral magnets with B20-type crystal structure such as MnSi~\cite{Muhlbauer09,Pappas09,Pfleiderer10,Adams11,Tonomura12}, Fe$_{1-x}$Co$_x$Si~\cite{YuXZ10N,Munzer10,Morikawa13}, FeGe~\cite{YuXZ10M}, and Mn$_{1-x}$Fe$_x$Ge~\cite{Shibata13} via small-angle neutron-scattering experiments and Lorentz transmission electron microscopies.

Subsequent intensive researches have revealed that magnetic skyrmions possess advantageous properties for application to high-density and low-energy-consuming storage devices~\cite{Fert13,Nagaosa13}, that is, (1) nanometric small size, (2) topological stability, (3) high transition temperatures, and (4) ultralow energy consumption to drive their motion. It was found that translational motion and subsequent Hall motion of skyrmions can be driven in metallic systems by applying spin-polarized electric currents via the spin transfer torque mechanism~\cite{Jonietz10,YuXZ12,Schulz12}. Surprisingly its threshold current density $j_{\rm c}$ turned out to be 10$^5$-10$^6$ A/m$^2$, which is five or six orders of magnitude smaller than $j_{\rm c}$ required to move domain walls.

The B20 compounds had been only example of chiral-lattice magnets realizing skyrmionic phases so far, and all of these compounds are metallic. The initial discovery of an insulating skyrmionic phase was reported in 2012 for Cu$_2$OSeO$_3$~\cite{Seki12a}. What's interesting here is that the noncollinear skyrmion spin structure in the insulator attains multiferroic nature via relativistic spin-orbit coupling.
Indeed, magnetically induced ferroelectric polarization was observed in this compound~\cite{Seki12a,Seki12b,Belesi12}. This multiferroicity offers an opportunity to manipulate skyrmions by electric fields~\cite{White12,White14} rather than electric currents~\cite{Everschor11,ZangJ11,Tchoe12,Iwasaki13a,Iwasaki13b,Sampaio13,SZLin13} or heat pulses~\cite{Koshibae14}. Because electric fields in insulators do not bring about energy losses due to the Joule heating in contrast to electric currents in metals, there is a chance to further reduce the energy consumption in potential skyrmion-based storage devices.

To use multiferroic skyrmions as information carriers, it is necessary to establish a method to create, erase, and drive them by applying an electric field. In this paper, by taking Cu$_2$OSeO$_3$ as an example of skyrmion-hosting multiferroics, we theoretically demonstrate that skyrmions can be created on a thin-film sample very quickly (within a few nanoseconds) by applying an electric field with an electrode tip. This electric activity of skyrmions turns out to be mediated by the magnetoelectric coupling between the swirling skyrmion spins and the electric polarizations in multiferroics, which is distinct in microscopic mechanism from the spin transfer torque as a major channel of electric control of magnetism in metallic magnets. Therefore this finding will lead to a unique technique for using multiferroic skyrmions for future skyrmion-based memory devices.

\begin{figure}
\includegraphics[scale=1.0]{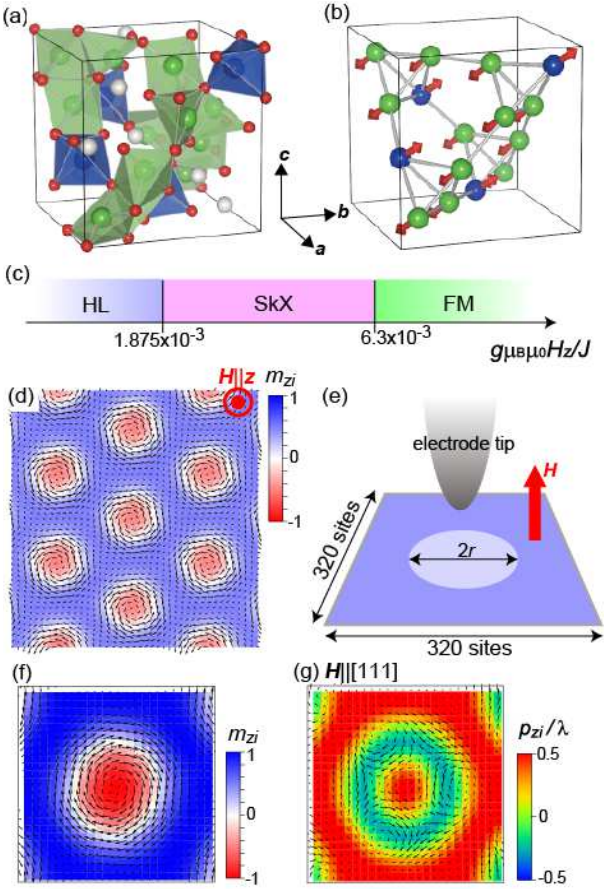}
\caption{(color online). (a) Chiral crystal structure of Cu$_2$OSeO$_3$ with P2$_1$3 symmetry. (b) Magnetic structure of Cu$_2$OSeO$_3$ with tetrahedra of four Cu$^{2+}$ ions ($S$=1/2) with three-up and one-down spins. (c) Phase diagram of the spin model~(\ref{eqn:model}). Here, HL, SkX, and FM denote helimagnetic, skyrmion-crystal, and ferromagnetic phases, respectively. (d) Spin structure of the hexagonal skyrmion crystal. In-plane magnetization components are represented by arrows. (e) Setup for the numerical simulations. A static magnetic field $\bm H$ is applied to the square lattice with 320$\times$320 sites in the perpendicular direction, while the static electric field $\bm E$ perpendicular to the plane is applied within a circular area whose diameter is $2r$ sites. (f) Real-space magnetization configuration of a skyrmion where the in-plane (out-of-plane) magnetization components are represents by arrows (colors). (g) Real-space map of local electric polarizations $\bm p_i$ in a skyrmion under $\bm H$$\parallel$[111]. Sum of the local contributions $\bm p_i$ on the {111} plane gives a finite net component $\bm P$$\parallel$[111].}
\label{Fig1}
\end{figure}
The crystal and magnetic structures of Cu$_2$OSeO$_3$ are composed of tetrahedra with four Cu$^{2+}$ ($S$=1/2) ions as shown in Figs.~\ref{Fig1}(a) and (b), and three-up and one-down type collinear spin arrangement is realized on each tetrahedron below $T_{\rm c}$$\sim$58 K~\cite{Bos08,Belesi10}. This four-spin assembly as a magnetic unit can be treated as a classical unit-vector magnetization $\bm m_i$. The magnetism in a thin-plate specimen of Cu$_2$OSeO$_3$ is described by a classical Heisenberg model on the square lattice~\cite{Bak80,YiSD09,HanJH10}. The Hamiltonian is given by,
\begin{eqnarray}
\mathcal{H}_0&=&
-J \sum_{<i,j>} \bm m_i \cdot \bm m_j
-D \sum_{i,\hat{\bm \gamma}} 
\bm m_i \times \bm m_{i+\hat{\bm \gamma}} \cdot \hat{\bm \gamma} 
\nonumber \\
&-&g\mu_{\rm B}\mu_0 H_z \sum_i m_{iz},
\label{eqn:model}
\end{eqnarray}
where $g$=2, and $\hat{\bm \gamma}$ runs over $\hat{\bm x}$ and $\hat{\bm y}$. Here, the $z$-axis is defnied as $\bm z$$\parallel$[111] with respect to the cubic $a$, $b$, and $c$ axis. The magnetization vector $\bm m_i$ is defined as $\bm m_i=-\bm S_i/\hbar$ in the units of m$^{-3}$ with $\bm S_i$ being the spin. The Hamiltonian contains the ferromagnetic-exchange interaction, the DMI~\cite{YangJH12} and the Zeeman coupling to the static magnetic field $\bm H$=(0,0,$H_z$) applied perpendicular to the plane. We use $J$=3 meV and $D/J$=0.09 so as to reproduce the magnetic transition temperature ($\sim$60 K) for bulk samples of Cu$_2$OSeO$_3$ and the skyrmion size ($\sim$50 nm) observed in a thin-plate specimen~\cite{Seki12a}.

The phase diagram of this model at $T=0$ is shown in Fig.~\ref{Fig1}(c). The skyrmion crystal phase emerges in the range $1.875\times10^{-3}<|g\mu_{\rm B}\mu_0 H_z/J|<6.3\times10^{-3}$, sandwiched by the helical and the ferromagnetic phases in agreement with experiments for a thin-plate sample of Cu$_2$OSeO$_3$~\cite{Seki12a}. Note that the ferromagnetic order of $\bm m_i$ corresponds to the ferrimagnetic order in real Cu$_2$OSeO$_3$ material because $\bm m_i$ represents the ferrimagentic three-up and one-down spin assembly. In the skyrmion-crystal phase, skyrmions are crystallized into a hexagonal lattice as shown in Fig.~\ref{Fig1}(d)~\cite{Seki12a,Adams12}, in which the magnetizations $\bm m_i$ point parallel (antiparallel) to $\bm H$ at the periphery (center) of each skyrmion. The phase transition between the skyrmion-crystal and ferromagnetic phases is of the strong first order, and thus skyrmions appear not only as a crystallized form but also as topological defects in the ferromagnetic phase. In the following, we demonstrate that isolated skyrmions can be created by applying an electric field with an electrode tip on a thin-film specimen in the ferromagnetic phase as shown in Fig.~\ref{Fig1}(e).

It was experimentally confirmed that the magnetizations in the noncollinear skyrmion structure induces electric polarizations $\bm p_i$ via the spin-dependent metal-ligand hybridization mechanism~\cite{Seki12b}. Because of the cubic crystal symmetry, the local polarization $\bm p_i$ in the units of Cm$^{-2}$ from the $i$th tetrahedron is given using the magnetization components $m_{ia}$, $m_{ib}$, and $m_{ic}$ in the units of m$^{-3}$ in the cubic setting as,
\begin{eqnarray}
\bm p_i=\left(p_{ia}, p_{ib}, p_{ic} \right)
= \lambda \left(m_{ib}m_{ic}, m_{ic}m_{ia}, m_{ia}m_{ib} \right).
\end{eqnarray}
The constant $\lambda$ is evaluated as $\lambda$=$5.64\times10^{-33}$ Cm from the experimental data~\cite{Belesi12,Miller10}. Spatial distributions of $\bm p_i$ induced by the skyrmion magnetizations $\bm m_i$ [Fig~\ref{Fig1}(f)] can be calculated from this equation, which vary depending on a choice of the thin-film plane. Shown in Fig.~\ref{Fig1}(g) is a real-space map of $\bm p_i$ on the {111} plane. We have confirmed that incorporation of electric dipole-dipole interactions never affects the electric-polarization distribution, indicating negligible roles of depolarization fields even in thin-film samples. This is because the electric polarization is a subsequent order parameter in this multiferroic system, which is governed by the predominant skyrmion magnetic order determined by strong magnetic interactions such as the ferromagnetic-exchange and the Dzyaloshinskii-Moriya interactions.

The net magnetization $\bm M$ and the ferroelectric polarization $\bm P$ are given by sums of the local contributions as $\bm M$=$\frac{g\mu_{\rm B}}{NV}\sum_{i=1}^{N} \bm m_i$ and $\bm P$=$\frac{1}{NV}\sum_{i=1}^{N} \bm p_i$, respectively. Here the index $i$ runs over the Cu-ion tetrahedra with three-up and one-down spin pair, $N$ is the number of the tetrahedra, and $V$(=1.76$\times$10$^{-28}$ m$^3$) is a volume per tetrahedron.

The coupling between magnetism and electricity offers an opportunity to create and manipulate magnetic skyrmions electrically through modulating the distribution of electric polarizations. To see this, we numerically simulate dynamics of magnetizations $\bm m_i$ and polarizations $\bm p_i$ under a locally applied electric field by solving the Landau-Lifshitz-Gilbert equation using the fourth-order Runge-Kutta method. The equation is given by
\begin{equation}
\frac{d\bm m_i}{dt}=-\bm m_i \times \bm H^{\rm eff}_i 
+\frac{\alpha_{\rm G}}{m} \bm m_i \times \frac{d\bm m_i}{dt},
\label{eq:LLGEQ}
\end{equation} 
where $\alpha_{\rm G}$(=0.04) is the Gilbert-damping coefficient. The effective field $\bm H_i^{\rm eff}$ is calculated from the Hamiltonian $\mathcal{H}$=$\mathcal{H}_0$+$\mathcal{H}'(t)$ as
$\bm H^{\rm eff}_i = -\partial \mathcal{H} / \partial \bm m_i$.
Here the first term $\mathcal{H}_0$ is the model Hamiltonian~(\ref{eqn:model}), while the term $\mathcal{H}'(t)$ represents the coupling between the local polarizations $\bm p_i$ and the dc electric field $\bm E$. The term $\mathcal{H}'(t)$ is given by
\begin{equation}
\mathcal{H}'(t)= -\bm E(t) \cdot \sum_{i \in \mathcal{C}} \bm p_i,
\end{equation}
where the electric field $\bm E(t)$=(0,0,$E_z$) is applied for a fixed time to the sites within a circular area $\mathcal{C}$ with diameter of $2r$=40 sites. Here, we model the local $\bm E$-field application as a constant $E_z$ within the area but $E_z$=0 otherwise. The calculations are performed using a system of $N$=320$\times$320 sites with an open boundary condition.

\begin{figure}
\includegraphics[scale=1.0]{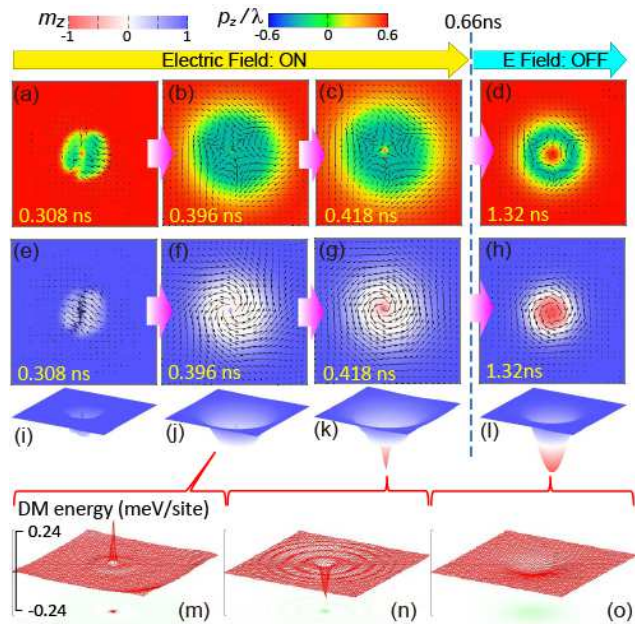}
\caption{(color online). (a)-(l) Snapshots of simulated spatiotemporal dynamics of polarizations $\bm p_i$ (a)-(d) and magnetizations $\bm m_i$ (e)-(l). In-plane (out-of-plane) components are represented by arrows (colors). The simulation is performed for a system of 320$\times$320 sites, while a relevant area of 100$\times$100 sites is magnified. The $\bm E$-field is applied in a circular area with diameter of $2r$=40 sites located in the middle of the system. (m)-(o) Snapshots of spatial distribution of energies associated with the Dzyaloshinskii-Moriya interaction.}
\label{Fig2}
\end{figure}
Our simulation demonstrates that isolated skyrmions can be created in the field-polarized ferromagnetic state under $\bm H$$\parallel$$[111]$. Simulated dynamics of $\bm m_i$ and $\bm p_i$ during the creation process are summarized in Figs.~\ref{Fig2}(a)-(l) for $g\mu_{\rm B}\mu_0 H_z/J$=$6.3\times10^{-3}$ and $E_z$=$-1.2\times10^{10}$ V/m. Relevant areas of 100$\times$100 sites are magnified in these figures. Note that although the value of $E_z$ is quite large, resulting effective magnetic fields due to the magnetoelectric coupling acting on $\bm m_i$ are not so large because of the small coupling constant $\lambda$, which allows us to neglect amplitude fluctuations of $\bm m_i$ in the simulation.

The application of $\bm E$ with negative $E_z$ induces reorientation of the polarizations $\bm p_i$ in the field-applied area from $\bm p_z$$>$0 to $\bm p_z$$<$0 as seen in Figs.~\ref{Fig2}(a)-(c). Accompanied with this $\bm p_i$ reorientation, most of the magnetizations $\bm m_i$ in the area rotate from the out-of-plane direction to the in-plane direction as seen in Figs.~\ref{Fig2}(e)-(g) and in Figs.~\ref{Fig2}(i)-(k). We find that a sudden 180$^{\circ}$ flop of local $\bm m_i$ occurs at the center of the field-applied area between (f) and (g) and between (j) and (k). Shown in Figs.~\ref{Fig2}(m)-(o) are snapshots of the spatial distribution of energy associated with the DMI. Importantly the energy becomes significantly high right before the $\bm m_i$ flop at a local site, and then suddenly becomes significantly low right after the $\bm m_i$ flop as exemplified by sharp positive and negative peaks in Figs.~\ref{Fig2}(m) and (n), respectively. Once the local $\bm m_i$ flop occurs, a skyrmion structure emerges after switching off the $\bm E$ field through relaxation of the spatial distributions of $\bm m_i$, $\bm p_i$ and the local Dzyaloshinskii-Moriya energy as shown in Figs~\ref{Fig2}(d), (h), (l) and (o). Whole process of this skyrmion creation occurs very quickly only within a few nanoseconds.

\begin{figure}[t]
\includegraphics[scale=1.0]{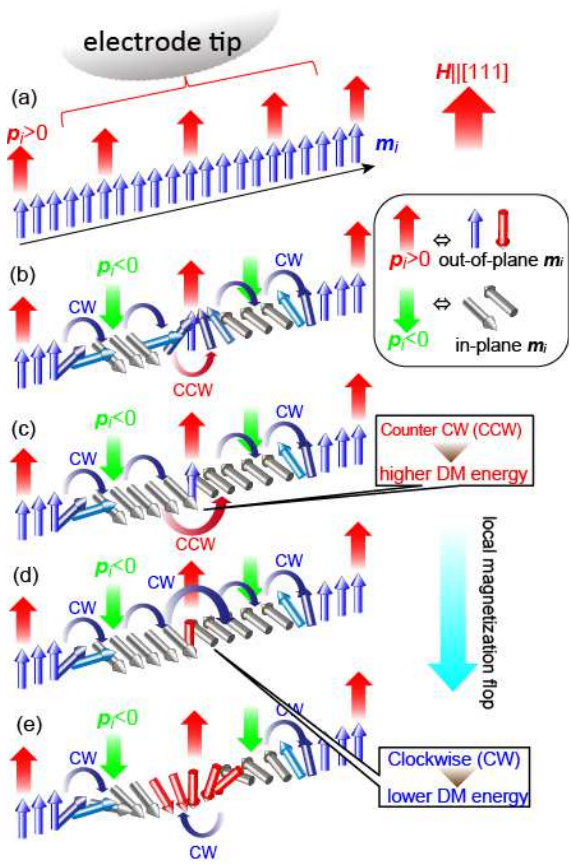}
\caption{(color online). Schematics of dynamical process of the electric-field-induced skyrmion formation, which focus on alignments of $\bm m_i$ and $\bm p_i$ in a diameter direction of the area onto which a dc electric field is applied.}
\label{Fig3}
\end{figure}
In order to understand the mechanism of this electrical creation of skyrmion, the following two facts should be noted. First the local $\bm m_i$ with dominant out-of-plane (in-plane) component or $\bm m_i$$\parallel$$[111]$ ($\bm m_i$$\perp$$[111]$) gives rise to local $\bm p_i$ with $p_z$$<$0 ($p_z$$>$0) under $\bm H$$\parallel$$[111]$ as seen in comparison between Figs.~\ref{Fig1}(f) and (g). Second the DMI in the model~(\ref{eqn:model}) with a positive parameter $D$$>$0 favors a clockwise rotation of $\bm m_i$ propagating in a positive direction indicated by a solid arrow in Fig.~\ref{Fig3}(a). 

Shown in Figs.~\ref{Fig3}(a)-(e) are schematics of the spatiotemporal dynamics of $\bm m_i$ and $\bm p_i$ aligned along a diameter of the electric-field area. In the initial ferromagnetic state with all the $\bm m_i$ being parallel to $\bm H$$\parallel$$[111]$, all the $\bm p_i$ are uniformly pointing in the $[111]$ direction with $p_z$$>$0 [Fig.~\ref{Fig3}(a)]. When the $\bm E$ field with $E_z$$<$0 is applied, the magnetizations $\bm m_i$ begin to rotate toward the in-plane direction so as to flop the $\bm p_i$ from $p_z$$>$0 to $p_z$$<$0 because $E_z$$<$0 favors $p_z$$<$0 [Fig.~\ref{Fig3}(b)]. This rotation of $\bm m_i$ occurs in the clockwise fashion near the periphery of the electric-field area so as to smoothly connect the spatial variation of $\bm m_i$ to the outside ferromagnetic region in the presence of the DMI. In this situation, the rotation sense around the center of the electric-field area inevitably becomes counterclockwise, which is unfavorable with respect to DMI. When the area of $p_z$$<$0 with in-plane $\bm m_i$ spreads under $E_z$$<$0, the $\bm m_i$ around the center become to rotate very abruptly in the counterclockwise fashion [Fig.~\ref{Fig3}(c)], which causes significantly high energy cost of the DMI as seen in the sharp positive peak in Fig.~\ref{Fig2}(m). To solve this energetically unstable alignment of $\bm m_i$, the $\bm m_i$ at the center eventually flops from $m_z$$>$0 to $m_z$$<$0 [Fig.~\ref{Fig3}(d)]. This local $\bm m_i$ flop realizes an abrupt clockwise rotation of $\bm m_i$ at the center with a large energy gain from the DMI seen as a sharp negative peak in Fig.~\ref{Fig2}(n). This locally inverted $\bm m_i$ becomes a skyrmion core. After switching off the $\bm E$ field [Fig.~\ref{Fig3}(d)], the alignment of $\bm m_i$ relaxes to form a skyrmion spin texture shown in Figs.~\ref{Fig2}(l)-(o).

Here, we argue some issues on the initial process of skyrmion creation. First, the signatures in Figs.~\ref{Fig2}(a) and (e) are not circular because the initial process is governed by a nucleation. A place at which the $\bm m_i$ begins to rotate is accidentally determined by fluctuations of $\bm m_i$ or imperfectness of skyrmion vortex structures in real experimental situations. This effect is incorporated via discreteness of the magnetization distribution or rounding errors in the numerical simulation. Second, the rotation of $\bm m_i$ under the $\bm E$-field first starts at peripheries, whereas the $\bm m_i$ at the center of the $\bm E$-field spot remains to point in the [111] direction. This is because the clockwise and counterclockwise rotations of $\bm m_i$ at the center surrounded by ferromagnetic background are degenerate so that it cannot start to rotate. On the other hand, at the peripheries of the $\bm E$-field spot, this degeneracy is lifted by asymmetry between inside and outside of the E-field area as well as the Dzyaloshinskii--Moriya interaction. There, the $\bm m_i$ can start to rotate in a direction favored by the Dzyaloshinskii--Moriya interaction. Third, although the numerical simulations are performed for a purely two-dimensional model, the skyrmion-creation process and the physical mechanism argued here are expected to survive even for film- or plate-shaped samples with finite thickness. There, the reversal of local $\bm m_i$ should occur first at the top layer, and subsequently the $\bm m_i$-reversed area expands in the thickness direction to form a tube-shaped skyrmion.

\begin{figure}
\includegraphics[scale=1.0]{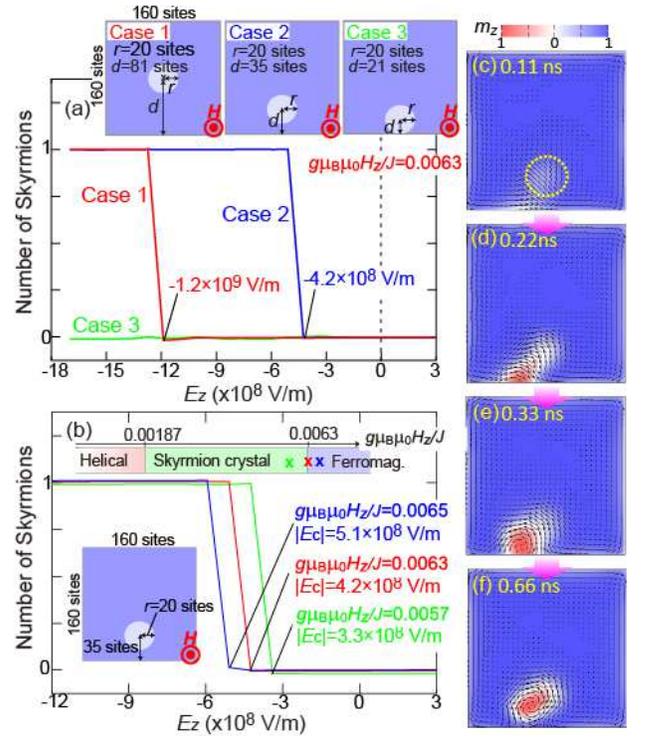}
\caption{(color online). (a), (b) $E_z$-dependence of the skyrmion creation for several different locations of the electric-field area (a) and for different values of external magnetic field $H_z$ (b). (c)-(f) Snapshots of the $\bm m_i$ dynamics for the skyrmion creation process when the $\bm E$ field ($E_z=-2.264 \times 10^8$ V/m) is applied in the dashed-circled area near the sample edge (Case 2). All the calculations are performed for a system of 160$\times$160 sites with an open boundary condition.}
\label{Fig4}
\end{figure}
We also discuss the $E$-field strength required for creating a skyrmion. Although efforts to search for new materials hosting multiferroic skyrmions with larger electric polarizations or stronger magnetoelectric coupling should be unremittingly made, it is useful to establish a method to reduce the threshold field for technical application. One promising way is to utilize a sample edge. To create a topological skyrmion texture deep inside a sample, one needs to flop the local $\bm m_i$ with a large energy cost. However, the $\bm m_i$ flop can be achieved with a much smaller energy at the sample edge. This is not only because the number of surrounding spins at the edge is small but also because discontinuity of the magnetization distribution at the sample edge allows continuous change of the topological invariant and relaxes the constraint of topological protection. 

In Fig.~\ref{Fig4}(a), the number of created skyrmions (0 or 1) is plotted as a function of $E_z$ for different locations of the electric-field area indicated by circles in the insets. When the $\bm E$ field is applied deep inside the sample (Case 1), the threshold field takes a negatively large value of $-1.2 \times 10^9$ V/m. On the other hand, application of the $\bm E$ field near the sample edge (Case 2) significantly reduces the threshold field to $-4.2 \times 10^8$ V/m. Snapshots of the dynamical $\bm m_i$ configuration for Case 2 are shown in Figs.~\ref{Fig4}(c)-(f), in which the $\bm m_i$ flop occurs at the sample edge. However, if the electric-field area is too close to the sample edge (Case 3), skyrmions cannot be created. More concretely although the local flop of $\bm m_i$ occurs at the edge with much lower energy in Case 3, the seed of skyrmion is absorbed by the edge and vanishes immediately after the $\bm E$ field is turned off. Note that typical threshold $E$ field for the dielectric breakdown is $\sim$100 MV/m if the $E$ field is applied to bulk of the sample with a plate electrode, but the dielectric breakdown does not occur even with a stronger field of 1-10 GV/m if the $E$ field is applied only locally with an electrode tip.

Another way to reduce the threshold field is tuning the external magnetic field $H_z$. The phase transition from the ferromagnetic phase to the skyrmion-crystal phase is of strong first order. As a result, the ferromagnetic phase remains as a metastable state even below the critical magnetic field. Figure~\ref{Fig4} shows $E_z$-dependence of the skyrmion creation for several values of $H_z$. We find that the threshold electric field is reduced as $H_z$ is decreased.

In summary, we have theoretically demonstrated that skyrmion spin textures can be electrically created on a thin-film specimen of chiral-lattice insulating magnet Cu$_2$OSeO$_3$ within a few nanoseconds by applying a dc electric field $\bm E$ with an electrode tip. The applied $\bm E$ field induces twisting of the magnetization alignment and eventually a 180$^{\circ}$ flop of the local magnetization through modulating the spatial distribution of DMI energy as well as the local polarization orientation via magnetoelectric coupling. The skyrmion spin structure grows around the flopped magnetization after switching off the $\bm E$ field. It was found that the required strength of $\bm E$ field can be significantly reduced by applying the $\bm E$ field onto the sample edge where the discontinuity of spatial $\bm m_i$ distribution at the sample edge relaxes the constraint from the topological protection. Recently, several theoretical proposals on how to drive skyrmions in insulators have been made~\cite{Mochizuki14,KongL13,LinSZ14a}, and experimental techniques to write skyrmions by STM tip with spin-polarized electric currents have been developed~\cite{Heinze11,Romming13}. Under these circumstances, control and creation of multiferroic skyrmions by electric fields promise to become an important technique towards spintronics application to realize low-energy-consuming storage devices.

The author would like to thank A. Rosch and Y. Watanabe for enlightening discussions.
This research was in part supported by JSPS KAKENHI (Grant Numbers 25870169 and 25287088).

\end{document}